\documentclass[cha,
 aip,
 amsmath,amssymb,
 reprint,%
floatfix,
]{revtex4-1}
\usepackage{graphicx}
\usepackage{dcolumn}
\usepackage{bm}
\usepackage[mathlines]{lineno}
\usepackage[utf8]{inputenc}
\usepackage[T1]{fontenc}
\DeclareUnicodeCharacter{0394}{$\Delta$}
\usepackage{mathptmx}
\usepackage{etoolbox}
\usepackage[table]{xcolor}
\usepackage{booktabs}
\usepackage{url}
\usepackage{hyperref}
\usepackage{multirow}
\usepackage{tabularx}
\newcolumntype{Y}{>{\centering\arraybackslash}X}

\makeatletter
\def\@email#1#2{%
 \endgroup
 \patchcmd{\titleblock@produce}
  {\frontmatter@RRAPformat}
  {\frontmatter@RRAPformat{\produce@RRAP{*#1\href{mailto:#2}{#2}}}\frontmatter@RRAPformat}
  {}{}
}%
\makeatother
\makeatletter
\let\selectlanguage\@gobble
\makeatother
\begin{document}
\title{Site-Selective Functional Group Classification of Auger-Electron Spectra and Core-Electron Binding Energies with Convolutional Neural Networks}
\author{Adam E. A. Fouda*}
\affiliation{Data Science Institute, The University of Chicago, Chicago, IL, 60637, United States}
\email{adamfouda@uchicago.edu}

\date{\today}
\begin{abstract}
X-ray spectroscopy techniques probe chemical states in systems with atom-site specificity; however, the site-selective characterization of local bond environments in novel research materials often relies on a combination of reference spectra, multiple complementary spectroscopic techniques and electronic structure calculations. Auger-electron spectroscopy has long accompanied x-ray photoelectron spectroscopy as a second modality to resolve chemical states with overlapping core-electron binding energies. Here I show that the wealth of information encoded in the Auger spectrum offers the opportunity to train convolutional neural networks for site-selective functional group classification in organic molecules directly from the Auger spectrum lineshape. Furthermore, the inclusion of the core-electron binding energy as an additional modality improves the classification, either by augmenting the input fitted intensities with the binding energy or by conditioning the classification on the binding energy via feature-wise linear modulation layers. The latter approach was previously developed for visual reasoning in image classification, and the present results show that this is the more robust approach for including the binding energy. This work demonstrates the potential for new data-driven characterization capabilities in spectroscopic techniques that are information-rich but difficult to interpret.
\end{abstract}
\maketitle

\section{Introduction}

The local bonding environment around specific atom sites can significantly alter the activity of functional materials and molecules in applications such as catalysis\cite{https://doi.org/10.1002/aenm.202200716,LIU2025124655} and organic synthesis\cite{Jurczyk2022,https://doi.org/10.1002/ejoc.202501188}. Spectroscopic techniques are used to characterize chemical systems. Infrared (IR) and Raman spectroscopy excite the vibrational modes between atoms and are widely used for bond-type identification. However, IR and Raman are global probes of the system's bonding environment, and higher-energy x-rays are required to interrogate the bonding environment around a specific atom site. X-rays excite inner-shell electrons tightly bound to an atom's nucleus, and the binding energies of the electrons in these core orbitals are affected by the target atom's local bond environment, an effect known as the chemical shift.\cite{bagus_mechanisms_1999} This phenomenon enables the photoionization of inner-shell electrons by x-ray photoelectron spectroscopy (XPS) measurements to resolve different chemical states in materials,\cite{bagus_interpretation_2013} molecules,\cite{siegbahn_esca_1970} solutions,\cite{seidel_valence_2016} liquids,\cite{hurisso_amino_2011,dick_probing_2020} and biological matter,\cite{ratner_surface_1983} with atom-site selectivity. However, the influence of multiple competing mechanisms, such as charge transfer, electric fields and hybridization,\cite{bagus_mechanisms_1999} which contribute to the structural environment effects on the core-electron binding energy, results in the presence of overlapping peaks in the XP spectrum, which limits the resolution of XPS measurements. The site-selective characterization of bond environments in novel research materials therefore often relies on reliable reference spectra and on the combination of multiple complementary spectroscopic techniques.

Following the removal of a core electron by x-ray photoionization, the core-hole decay process results in the autoionization of secondary electrons from the x-ray-irradiated sample.\cite{Siegbahn1967ESCAA,10.1021/ac60314a015} The detection of these secondary electrons, which appear as Auger features in the XP spectrum, is commonly used to distinguish oxidation states of active metal sites in catalysts and functional materials.\cite{doi:10.1021/acsanm.5c00100,https://doi.org/10.1002/sia.6239,SOLDEMO2024122565,HENDERSON2025147578,FOX1977390,10.1039/c2cp22419d,https://doi.org/10.1002/admi.202201828,doi:10.1021/jp0564400,doi:10.1021/acsomega.1c05002} This decay process, known as Auger-Meitner decay, involves an outer-shell electron filling the core vacancy, and another outer-shell electron being ejected to the continuum. All energetically possible decay channels involving two outer-shell electrons contribute to the spectrum, and the number of possible final states thus scales non-linearly with respect to the system size. The multitude of decay channels explicitly involving transitions of electrons from valence bonds to the core-hole encodes a wealth of information into the Auger spectrum, making it a sensitive probe of the surrounding electronic structure. Auger-Electron Spectroscopy (AES) is therefore a powerful tool for characterizing local bond environments in novel chemical systems. 

The complexity of the Auger-Meitner decay process enables its enhanced sensitivity compared with XPS and other single-electron x-ray techniques. However, its complexity also challenges its experimental interpretation and computational simulation, which limits the widespread adoption of this technique. The theoretical treatment of the large number of final valence two-hole states (including the contribution of multielectron shakeup and shakeoff processes) is non-trivial due to their multiconfigurational character.\cite{doi:10.1063/1.1386414,10.1063/5.0062130,doi:10.1021/acs.jpca.5c01789} Furthermore, the accurate determination of the Auger-Meitner decay rate, and thus spectral intensity, requires the treatment of the ejected electron's continuum wave function. Numerous approaches have been developed for its treatment, including approximating the spectral intensity by an electron population analysis,\cite{MITANI2003103,10.1063/1.2166234,D3CP01746J,doi:10.1021/acs.jpclett.3c03611} and both implicit\cite{10.1063/1.1316046,BSchimmelpfennig_1992,SCHIMMELPFENNIG1995173,LIEGENER1982188,10.1063/1.2126976,10.1063/5.0036976,SIEGBAHN1975330,JENNISON1980435,Larkins1990,FINK1995295,TRAVNIKOVA200967,PhysRevA.94.023422,10.1063/1.4919794} and explicit\cite{PhysRevA.19.1649,HIGASHI1982377,PhysRevA.45.318,Demekhin2007,PhysRevA.80.063425,10.1063/1.3526026,10.1063/1.3700233,C7CP02345F} considerations of the continuum electron wave function. Thus, full treatments of the Auger spectrum are usually limited to small molecules, and approximate treatments of the decay channels and rates are often employed for molecules with more than 20 atoms and for non-gas-phase systems.

Whilst the analysis of x-ray spectroscopy measurements with electronic structure methods in ``\textit{forward}'' (structure-to-spectrum) mapping is well established for techniques such as XPS,\cite{dick_probing_2020,Fouda2017,10.1021/acs.jctc.6c00713,hirao_theoretical_2025-1,besley_density_2021} x-ray absorption spectroscopy (XAS),\cite{10.1039/c002207a,10.1063/1.2967190,ASMURUF2008267,10.1021/jp065160x,10.1039/b912718f,BUCKLEY2011540,10.1021/acs.jctc.6b00656} x-ray emission spectroscopy (XES)\cite{10.1021/ct500566k,https://doi.org/10.1002/jcc.26153,10.1021/acs.jpca.9b08037,FOERSTER2020137860,10.1063/1.4977178,BESLEY201242} and resonant inelastic x-ray scattering (RIXS),\cite{besley_density_2020,besley_modeling_2021,10.1021/acs.jctc.8b00211,fouda_observation_2020,nascimento_resonant_2021,rankine_progress_2021,10.1021/acs.jctc.6c00739} the recent rise in data-driven approaches utilizing machine learning (ML) has enabled rapid growth in developments using ``\textit{backward}'' (spectrum-to-structure) mapping in XPS,\cite{PIELSTICKER2023341433,Drera_2020,vakhrushev_application_2024,simperl_transformer_2025,de_curto_large_2024} XAS\cite{Guda2021,10.1021/acs.chemmater.3c02584,PhysRevMaterials.3.033604,10.1021/jacs.2c11824} and XES\cite{10.1039/d1cp02903g}. The information rich and computationally complex nature of AES makes this technique particularly well suited for ``\textit{backward}'' mapping. Prior to the recent rise in ML applications in the physical sciences, neural networks have been applied to the classification of single element Auger spectra\cite{https://doi.org/10.1002/sia.740201303}, discriminating titanium nitride compositions in thin films,\cite{https://doi.org/10.1002/sia.740230709} and background removal for automatic analysis\cite{https://doi.org/10.1002/sia.5011}. 

This work explores a novel data-driven approach for the site-specific identification of local bond environments in organic molecules (organic functional groups) directly from the Auger spectrum lineshape with a convolutional neural network (CNN), a class of model commonly used in computer vision applications. CNNs extract local patterns from high-dimensional inputs via several layers of convolution filters to form a hierarchical composition of increasingly complex features. One-dimensional (1D) CNNs have demonstrated success in the analysis of spectral signals IR and Raman spectroscopy,\cite{Fuentes2023,10.1039/d3dd00203a,10.1117/12.2618487} and XAS\cite{PhysRevMaterials.3.033604} For XPS, Drera \textit{et al.} used a CNN to quantify elemental stoichiometry from XP spectra,\cite{Drera_2020} whilst Pielsticker \textit{et al.} used a CNN to determine the concentration of different chemical phases in transition metal XP spectra.\cite{PIELSTICKER2023341433} Both studies consider the effect of the Auger features in the XP lineshapes, and Pielsticker \textit{et al.} found that correctly quantifying the different chemical phases in spectra with multiple elements required the consideration of Auger transitions. To the best of my knowledge no CNN has been previously applied to the analysis of molecular AES. 

In this work, I present a CNN model for classifying the organic functional groups directly from the carbon $1s$ Auger spectrum. The absence of large experimental databases in this domain necessitates model training on a dataset of calculated spectra. The spectra were calculated on a subset of molecules from the QM9 database using an electronic structure method which was previously benchmarked against experiment with modest accuracy.\cite{doi:10.1021/acs.jpca.5c01789} The level of theory and construction of the dataset are discussed in Section \ref{sec:data}. Furthermore, this work shows that the classification performance is improved by including the carbon $1s$ binding energy ($E_b$) as an additional input via two different approaches. The first approach is a simple augmentation of $E_b$ to the fitted Auger spectrum intensities. The second approach conditions the classification with the $E_b$ via the inclusion of feature-wise linear modulation (FiLM) layers, which apply a feature-wise affine transformation between the convolution layers. The implementation of FiLM layers into the CNN architecture is discussed in Section \ref{sec:model}. FiLM layers were developed to condition image classification with text inputs to achieve visual reasoning. I show that this mechanism can be translated to chemical applications and provides an new strategy for multi-modal chemical state identification in automated spectroscopy analysis.

\section{Computational Details}

\subsection{Dataset}\label{sec:data}

\begin{table}[htbp]
\centering
\caption{Carbon environment classes identified in the dataset with the SMARTS string pattern matching algorithm, with the merged class names in bold. A symbol representation of the nearest neighbor bonding defining the un-merged classes are given along side the counts of each class in the training and validation dataset and the hold-out and evaluation test sets. SM Table S1 contains the SMARTS strings\cite{daylight_chemical_information_systems_smarts-language_2019} used for the class definitions.}
\label{tab:carbon_env_counts}
\footnotesize
\setlength{\tabcolsep}{4pt}
\begin{tabular}{llrrr}
\toprule
Environment & Bonding & Train+val & Hold-out & Eval \\
\midrule
\multicolumn{2}{l}{\textbf{heteroaromatic}} & \textbf{1979} & \textbf{72} & \textbf{6} \\
arom N & N-\textbf{C}$_{\mathrm{ar}}$ &  1314 &    48 &     4 \\
arom O & O-\textbf{C}$_{\mathrm{ar}}$ &   354 &    17 &     1 \\
arom O N & O-\textbf{C}$_{\mathrm{ar}}$-N &   311 &     7 &     1 \\
\midrule
\multicolumn{2}{l}{\textbf{aryl N}} & \textbf{207} & \textbf{7} & \textbf{0} \\
aryl amine & \textbf{C}$_{\mathrm{ar}}$-NH$_{2}$ &   187 &     6 &     0 \\
aryl nitro & \textbf{C}$_{\mathrm{ar}}$-NO$_{2}$ &    20 &     1 &     0 \\
\midrule
\multicolumn{2}{l}{\textbf{aryl O}} & \textbf{240} & \textbf{7} & \textbf{0} \\
phenol & \textbf{C}$_{\mathrm{ar}}$-OH &   187 &     5 &     0 \\
aryl ether & \textbf{C}$_{\mathrm{ar}}$-O-R &    53 &     2 &     0 \\
\midrule
\multicolumn{2}{l}{\textbf{aryl F}} & \textbf{210} & \textbf{7} & \textbf{0} \\
aryl fluoride & \textbf{C}$_{\mathrm{ar}}$-F &   210 &     7 &     0 \\
\midrule
\multicolumn{2}{l}{\textbf{aryl carbonyl}} & \textbf{313} & \textbf{13} & \textbf{0} \\
aryl carbonyl & \textbf{C}$_{\mathrm{ar}}$=O &   313 &    13 &     0 \\
\midrule
\multicolumn{2}{l}{\textbf{hydrocarbon}} & \textbf{1603} & \textbf{32} & \textbf{31} \\
methyl & -\textbf{C}H$_{3}$ &   140 &     6 &    10 \\
methylene & -\textbf{C}H$_{2}$- &   170 &     5 &    10 \\
methine & $>$\textbf{C}H- &    65 &     2 &     0 \\
quaternary & $>$\textbf{C}$<$ &    14 &     0 &     0 \\
alkyne & -\textbf{C}$\equiv$C- &   774 &     4 &     2 \\
vinyl & \textbf{C}=C &    71 &     4 &     2 \\
aromatic & \textbf{C}$_{\mathrm{ar}}$ (all-C ring) &   369 &    11 &     7 \\
\midrule
\multicolumn{2}{l}{\textbf{carbonyl}} & \textbf{558} & \textbf{19} & \textbf{3} \\
ketone & R-\textbf{C}O-R &   148 &     3 &     1 \\
aldehyde & R-\textbf{C}HO &   229 &    10 &     2 \\
ester carbonyl & -\textbf{C}O-OR &   111 &     3 &     0 \\
ester alkyl & \textbf{C}-O-CO-R &    65 &     3 &     0 \\
carboxylic acid & -\textbf{C}OOH &     3 &     0 &     0 \\
carboxylate & -\textbf{C}OO$^{-}$ &     2 &     0 &     0 \\
\midrule
\multicolumn{2}{l}{\textbf{amide carbonyl}} & \textbf{167} & \textbf{3} & \textbf{1} \\
amide carbonyl & -\textbf{C}O-N &   165 &     3 &     1 \\
isocyanate & O=\textbf{C}=N &     2 &     0 &     0 \\
\midrule
\multicolumn{2}{l}{\textbf{nitrile}} & \textbf{447} & \textbf{12} & \textbf{0} \\
nitrile & -\textbf{C}$\equiv$N &   447 &    12 &     0 \\
\midrule
\multicolumn{2}{l}{\textbf{imine}} & \textbf{104} & \textbf{4} & \textbf{0} \\
imine & \textbf{C}=N &   104 &     4 &     0 \\
\midrule
\multicolumn{2}{l}{\textbf{oxyl}} & \textbf{274} & \textbf{3} & \textbf{0} \\
ether & \textbf{C}-O-C &   179 &     1 &     0 \\
alcohol & \textbf{C}-OH &    95 &     2 &     0 \\
\midrule
\multicolumn{2}{l}{\textbf{amine}} & \textbf{97} & \textbf{0} & \textbf{0} \\
amine & \textbf{C}-N &    97 &     0 &     0 \\
\midrule
\multicolumn{2}{l}{\textbf{alkyl fluorinated}} & \textbf{16} & \textbf{0} & \textbf{1} \\
fluorinated & \textbf{C}-F &    16 &     0 &     1 \\
\bottomrule
\end{tabular}
\end{table}

\begin{figure}[!htbp]
    \centering
    \includegraphics[width=1.0\columnwidth]{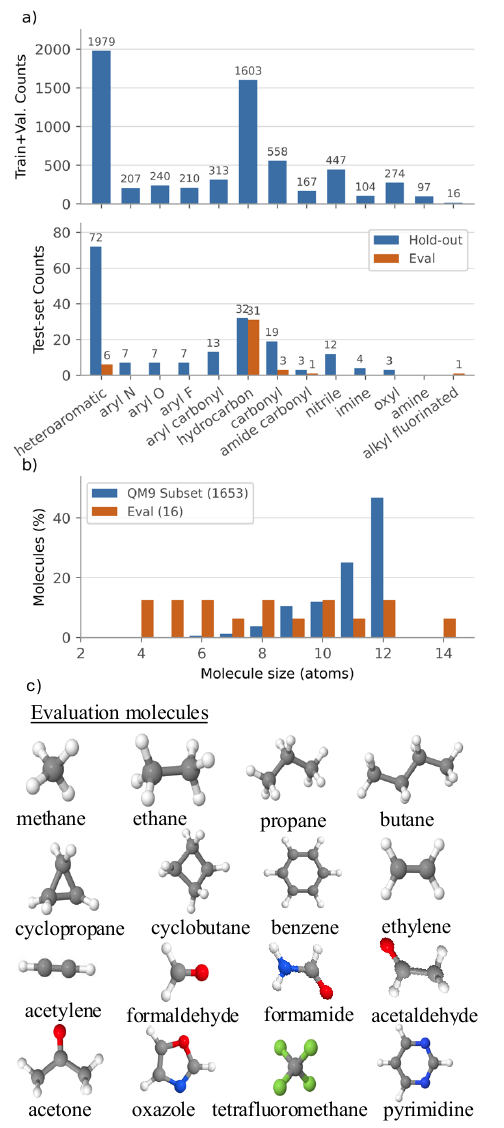}
    \caption{a) Bar charts of the merged environment class counts in the training and validation sets (upper panel) and hold-out (lower panel blue) and evaluation (lower panel dark orange) test sets. b) Percentage of molecular sizes in the QM9 subset (training, validation and hold-out dataset) (blue) and evaluation dataset (dark orange). c) Structures of the evaluation dataset molecules used in this work.}
    \label{fig:data}
\end{figure}

Recently, we showed that carbon 1$s$ Auger spectra in modest agreement with experiment can be produced efficiently for a set of 20 organic molecules by excluding the contribution of shakeup and shakeoff processes, implicitly treating the continuum with pre-calculated atomic bound-continuum integrals via the one-centre approximation (OCA),\cite{SIEGBAHN1975330,JENNISON1980435,Larkins1990,FINK1995295,TRAVNIKOVA200967,PhysRevA.94.023422,10.1063/1.4919794,tenorio_multi-reference_2021} and treating the bound-electron structure of the final dication states with the multiconfiguration pair-density functional theory\cite{mcpdft2014,mcpdftreview01,mcpdftreview02,mcpdftreview03} (MC-PDFT) method.\cite{doi:10.1021/acs.jpca.5c01789} MC-PDFT adds correlation effects through a straightforward computation from the one- and two-particle reduced density matrices and the optimized orbitals of a restricted active space self-consistent field (RASSCF) wave function.\cite{werner1981quadratically,malmqvist1990restricted} RASSCF enables the description of both core-hole and multiconfigurational valence dication states. The wave function used in this work excludes virtual ground-state orbitals, which enables high-throughput AES simulation but neglects shakeup and shakeoff contributions. The dataset of simulated carbon 1$s$ Auger spectra and $E_b$ values was computed from MC-PDFT electronic structure calculations with the tPBE0 functional and the ANO-RCC-VTZP basis set, combined with OCA decay rates,\cite{tenorio_multi-reference_2021} in the OpenMolcas software package.\cite{fdez2019openmolcas,li_manni_openmolcas_2023} 

Due to the computational costs of computing AES, the spectra were computed on a randomly selected subset of 1653 molecules, containing 3-12 atoms from the QM9 database.\cite{ramakrishnan_quantum_2014}This set of molecules are a subset of the molecules previously used to train equivariant graph neural network predictions of carbon 1$s$ $E_{b}$ values.\cite{fouda2026experimentallyaccurategraphneural} The number of final dication states for each molecule was limited to 300 singlet and 300 triplet states, which were combined and fitted to a kinetic energy grid of 751 points between 200 and 275 eV. Unless stated otherwise, a Gaussian FWHM of 1.6 eV was applied. Each individual carbon spectrum was normalized to its maximum intensity; the CNN therefore considers only the lineshape and omits the relative intensities between carbon environments. The $E_{b}$ values are represented as $\Delta E_{b}$ values ($E^{Atom}_{b}-E^{Mol}_{b}$) and were mean and standard deviation normalized using values determined across the set of training data carbons to avoid data leakage. The dataset contains 6394 individual carbon 1$s$ Auger spectra and 1$s$ $E_{b}$ values. The full data will be made publicly available.

The carbon environment classes corresponding to different organic molecule functional groups are defined by only the nearest neighbors to each carbon. They were determined by a SMARTS string\cite{daylight_chemical_information_systems_smarts-language_2019} pattern matching algorithm. The algorithm assigns a priority score to each environment class; these scores were used in cases where a carbon's nearest neighbors matched multiple possible environments. More constrained environments, such as carbonyl and aromatic environments, have a higher priority. Supplementary material (SM) Table S1 contains the environments, the corresponding SMARTS patterns, and the priority score. Thirty environment classes were identified in the dataset, and the assigned classes were validated against a similar procedure implemented in the Open Babel software package.\cite{OBoyle2011} The 30 classes were also merged into broader chemical classes to compare against the classification of coarser-grained definitions. Table \ref{tab:envs} lists the environment classes, with the merged classes in bold. Symbol representations for the un-merged classes are given alongside the counts of each class in the different dataset splits used in this work.

Initially, a randomly selected sample of 50 molecules was extracted from the QM9 subsample to form the hold-out test set. The remaining molecules then underwent 10-fold cross-validation (CV) using the Butina splitting approach\cite{butina_unsupervised_1999} to define the training and validation sets. In addition to the hold-out test, an evaluation test set is included which contains calculated spectra for 16 organic molecules whose experimental Auger spectra are available in the literature, and which were used in the study benchmarking the electronic structure calculations.\cite{doi:10.1021/acs.jpca.5c01789} This evaluation set represents a sample of the distribution of organic molecules found in AES studies in the literature. Figure \ref{fig:data} a) shows a bar chart of the merged environment class counts of the train and validation datasets (upper panel blue) and the test hold-out (lower panel blue) and evaluation (lower panel dark orange) datasets. The datasets derived from the QM9 subset (train, validation and hold-out) are imbalanced towards the merged heteroaromatic class. Table \ref{tab:envs} shows that this merged class is dominated by the nitrogen substituted aromatic ring class (arom N). The next dominant merged class is the hydrocarbon class, which represents un-merged classes with no non-hydrogen heteroatoms in the carbons nearest neighbors. Whilst the distribution of merged environments classes between the train, validation and hold-out sets is similar, the much small number of samples in the evaluation molecules strains any direct comparison of the class distributions with this dataset.

Figure \ref{fig:data} b) shows the distribution of molecular sizes between the QM9 subset and the evaluation molecules. The QM9 subset is dominated by 12-atom molecules, whilst the evaluation set has a higher proportion of smaller molecules containing between 4 and 8 atoms. Therefore, the evaluation dataset puts a spotlight on the model's capability to classify smaller systems and provides a measure of the model's size transferability. The structures of the evaluation molecules are shown in Figure \ref{fig:data} c).

\subsection{Convolutional Neural Network Architecture and Training}\label{sec:model}

\begin{figure}[!htbp]
    \centering
    \includegraphics[width=1.0\columnwidth]{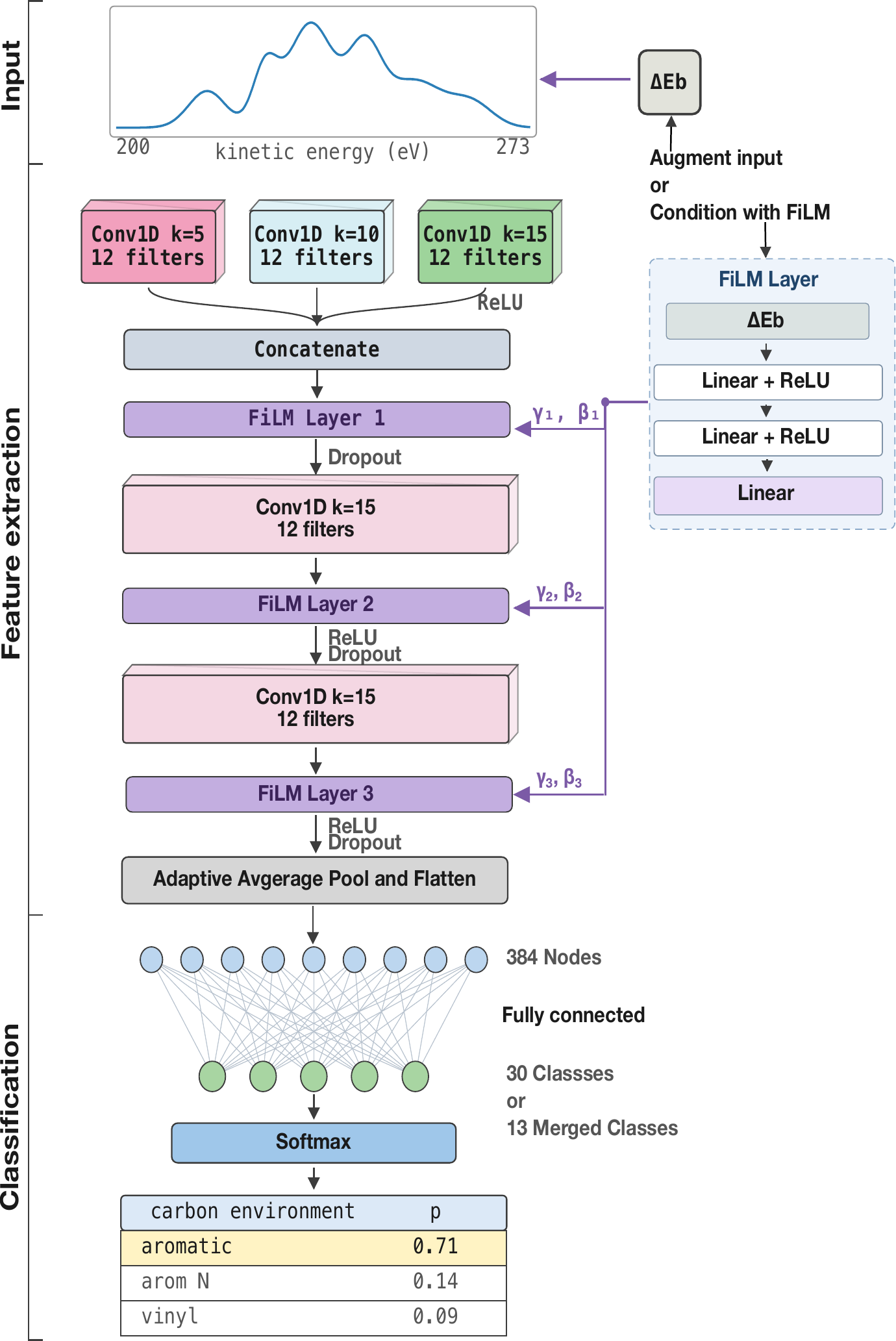}
    \caption{Schematic of the convolution neural network (CNN) architecture used for functional group classification from Auger spectra, with the inclusion of $E_b$ either through augmentation to the input fitted intensities or via conditioning with feature-wise linear modulation (FiLM) layers. The input is first passed through three parallel convolution layers each with 12 filter and kernel sizes of 5, 10 and 15 respectively. The outputs of parallel blocks under ReLU activation and are concatenated before being either passed through the first FiLM layer or passed straight to two sequential convolution layers. Two additional FiLM layers are applied between the sequential layers, both with 12 filters and kernel sizes of 15. The final step of the feature extraction is the an adaptive pool step which reduces the feature maps from 751 to 32. The pooled feature maps are then passed through a fully connected classifier, where the outputs are passed through SoftMax to produce class probabilities. Both the architectural, and schematic, design were inspired by previous work using a CNN to quantify chemical state compositions in transitions state XPS measurements.\cite{PIELSTICKER2023341433}}
    \label{fig:cnnfilm}
\end{figure}

\begin{figure*}[!htbp]
    \centering
    \includegraphics[width=1.0\textwidth]{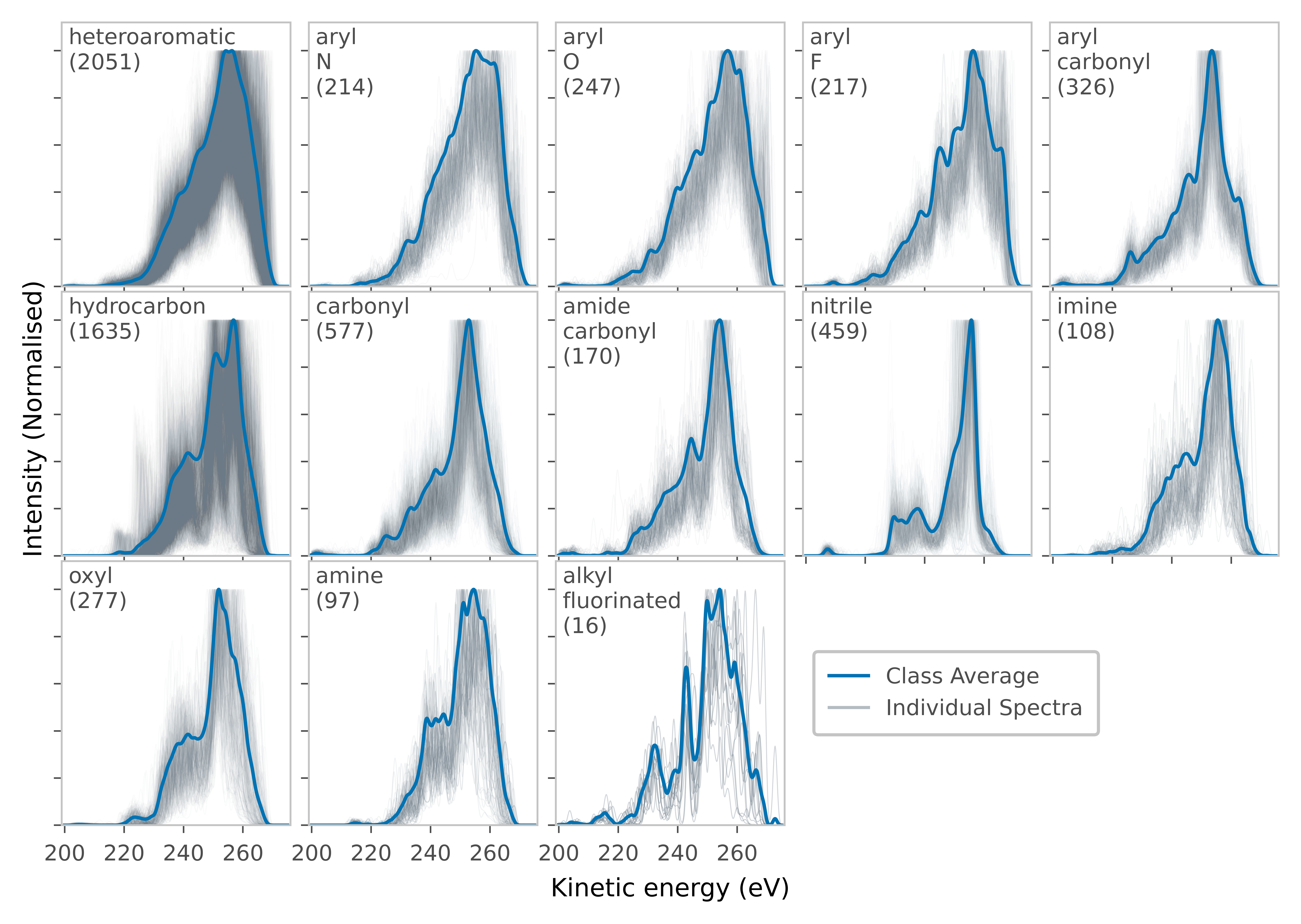}
    \caption{Lineshapes of the carbon 1$s$ Auger-spectra calculated on the full subset of QM9 molecules in this work. Each carbon spectrum (grey) is normalized to its intensity maximum. The spectra are shown across the 13 merged classes and the class average lineshape is shown (blue). SM Figure 1 contains the lineshapes across full set of 30 un-merged classes.}
    \label{fig:spectra}
\end{figure*}

\begin{figure*}[!htbp]
    \centering
    \includegraphics[width=1.0\textwidth]{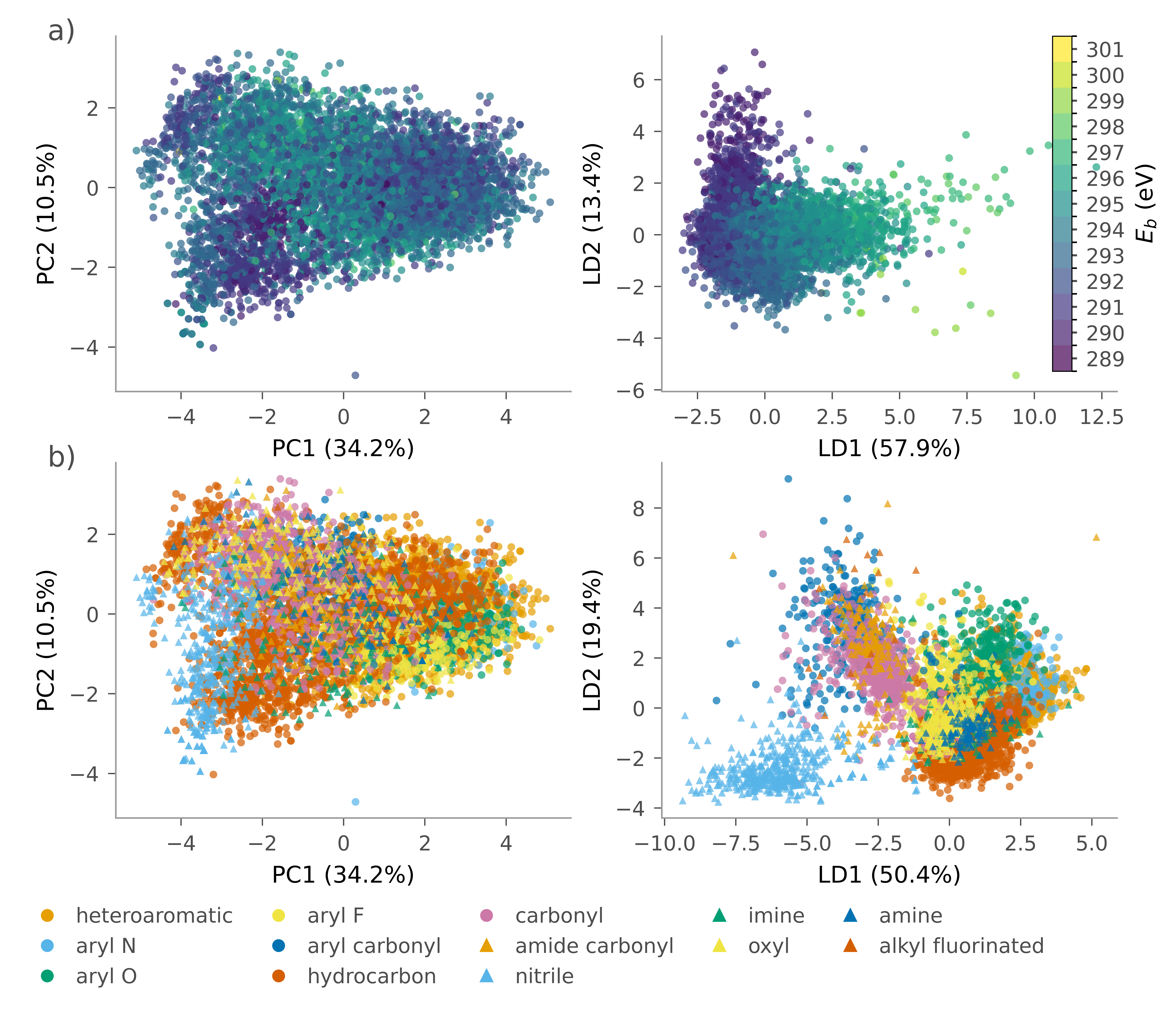}
    \caption{Principal component analysis (PCA) and linear discriminant analysis plots of lineshapes of the carbon 1$s$ Auger-spectra calculated on the full subset of QM9 molecules in this work. a) colors the scatter points by $E_b$ categories with $\pm0.5$ eV bins and b) colors and shapes the scatter points by the merged class definitions.}
    \label{fig:pcalda}
\end{figure*}

The architecture of the model used in this work takes inspiration from a previous study using a CNN to determine the concentration of different chemical phases within transition metal XPS measurements.\cite{PIELSTICKER2023341433} Pielsticker \textit{et al.} implemented a hybrid architecture comprised of a CNN feature extractor, followed by a fully connected quantifier. As the energy range, grid size and broadening widths of the Auger spectra in the present work are similar to the XP spectra from the previous study, the same kernel sizes and number of filters were also applied here. 

The key difference between the present architecture and the previous one is the inclusion of the FiLM layers, which condition the classification of the local bond environments from the Auger spectra with $\Delta E_{b}$ ($E^{Atom}_{b}-E^{Mol.}_{b}$, z-normed across the training data). The FiLM layers apply a linear modulation to the neural networks activations ($\textbf{\textit{F}}_{i,c}$) between the convolutional layers, with an affine transformation\cite{3504035.3504518}. This involves a learned multiplicative scaling and offset via,
\begin{equation}\label{eq:film}
FiLM(\textbf{\textit{F}}_{i,c}|\gamma_{i,c},\beta_{i,c}) = \gamma_{i,c}\textbf{\textit{F}}_{i,c} + \beta_{i,c}.
\end{equation}
Where subscripts denote the $c^{\mathrm{th}}$ feature map of the $i^{\mathrm{th}}$ input spectrum. $\gamma_{i,c}$ and $\beta_{i,c}$ are the outputs of arbitrary functions which receive the FiLM conditioning input $x_i$ ($\gamma_{i,c} = f_{c}(x_{i})$ and $\beta_{i,c} = h_{c}(x_{i})$). The present work only considers $\Delta E_{b}$ as a FiLM input. Following the original implementation,\cite{3504035.3504518} $f$ and $h$ are combined into a single FiLM generator, constructed by 3 fully connected layers with a hidden dimension of 64 and Rectified Linear Unit (ReLU) activation between the layers in this work. The output of the generator is split into the scale $\gamma_{c}$ and shift $\beta_{c}$ parameters, which modulate the features mapspassed through the network by Equation \ref{eq:film}. As both parameters are applied identically across all 751 kinetic energy points, the conditioning is agnostic to kinetic energy and re-weights the entire feature map as a function of the chemical shift. Therefore, the same convolutional blocks are interpreted differently for a carbon at a low $E_{b}$ than for one at a high $E_{b}$. In order to compare the FiLM conditioned classification to the un-conditioned classification from just the Auger spectrum, or the $\Delta E_{b}$ augmented Auger spectrum, the scale parameter is set to $\gamma_{c} = 1 + \Delta\gamma_{c}$ and the final layer of the generator is zero-initialized, so $\gamma_{c}=1$ and $\beta_{c}=0$ at the start of training. Which initiates the conditioned model to be identical with the unconditioned one.

Figure \ref{fig:cnnfilm} presents a schematic of the present architecture (the schematic design was also inspired by the study from Pielsticker \textit{et al.}\cite{PIELSTICKER2023341433}). The CNN input is a Gaussian fitted Auger spectrum across on grid of 751 points between 200 and 275 eV, optionally augmented with $\Delta E_{b}$ if FiLM conditioning is not used. The input is then passed to three parallel 1D convolution layers with kernel sizes $k= 5, 10, 15$ respectively, which consider spectral features of different sizes\cite{Drera_2020,PIELSTICKER2023341433} (0.5, 1.0 and 1.5 eV respectively). Each parallel convolution layers has 12 filters and a stride of 1, followed by ReLU activation and the concatenation of their outputs. Following the concatenation, the feature vectors are either passed through two sequential 1D convolutional layers with kernel sizes of 15 and 12 filters, or if FiLM conditioning is applied, the concatenated feature vector is passed through the FiLM layer before, in-between and after the two sequential convolutional layers. ReLU activation is applied after each sequential convolutional or FiLM layer. Dropout with rate 0.2 is applied after the concatenation or first FiLM layer, then after each ReLU activation in the sequential block. The final stage of the feature extraction is the pooling layer, which uses adaptive pooling with the number of outputs set to 32, to reduces the size of the features maps from 751 grid points to 32. 

This work swaps the previous quantifier for a classifier consisting of two fully connected layers, with an initial size of 384 (determined by the pooling output (32) and number of filters (12)) and the final layer size is the number of classes. Which is 30 for the un-merged bond environments and 13 for the merged bond environments. The final logits are then passed through SoftMax to yield the probabilities for each class. 

The AdamW optimizer was used for model training, with a learning rate of $3\times10^{-4}$, weight decay of $5\times10^{-4}$ and batch size was 64. The learning rate annealed by a cosine schedule over the full 500-epoch budget to a floor of $10^{-6}$. A cross-entropy loss function was used, with label smoothing of 0.1 and inverse-frequency class weights, normalized to unit mean over the populated classes. The inverse-frequency class weights compensates for class imbalances in the dataset (see Figure \ref{fig:data}). Early stopping of the training was applied if the macro-averaged $F_{1}$ of the validation set did not improve within 40 epochs. The reported model for each fold is taken from the epoch of highest validation macro $F_{1}$. The predicted classes were determined by the class with the largest probability.

\begin{table*}[t]
\centering
\renewcommand{\arraystretch}{1.3}
\caption{10 fols cross-validation (CV) avergaed results of theCNN classifications of organic functional groups fron the fitted Auger spectrum lineshape (Baseline) and models with the core-electron binding energy ($E_{b}$) included via augementation ($E_{b}$-aug)and FiLM layers  ($E_{b}$-FiLM). CV averaged accuracy, macro recall and macro F$_{1}$scores with the standard deviations are are given for training on the30 un-merged classes and 13 merged classes.}
\label{tab:cnn_cv}
\begin{tabular*}{\textwidth}{@{\extracolsep{\fill}}l|cc|cc|cc@{}}
\toprule
 Model & \multicolumn{2}{c|}{Accuracy (\%)} & \multicolumn{2}{c|}{Macro Recall} & \multicolumn{2}{c}{Macro $F_1$}\\
 & Hold-out & Eval-calc & Hold-out & Eval-calc & Hold-out & Eval-calc\\
\midrule
Un-Merged Baseline: & 76.6 $\pm$ 3.1 & 43.3 $\pm$ 9.9 & 0.70 $\pm$ 0.06 & 0.34 $\pm$ 0.07 & 0.67 $\pm$ 0.06 & 0.32 $\pm$ 0.06\\
\quad + $E_{b}$ aug. & 78.5 $\pm$ 2.6 & 47.6 $\pm$ 10.3 & 0.71 $\pm$ 0.05 & 0.37 $\pm$ 0.06 & 0.68 $\pm$ 0.05 & 0.34 $\pm$ 0.06\\
\quad + $E_{b}$ FiLM & 80.0 $\pm$ 1.2 & 53.6 $\pm$ 9.5 & 0.73 $\pm$ 0.03 & 0.53 $\pm$ 0.10 & 0.70 $\pm$ 0.03 & 0.48 $\pm$ 0.10\\
\midrule
Merged Baseline: & 80.2 $\pm$ 2.2 & 76.4 $\pm$ 4.0 & 0.69 $\pm$ 0.04 & 0.39 $\pm$ 0.10 & 0.68 $\pm$ 0.04 & 0.37 $\pm$ 0.10\\
\quad + $E_{b}$ aug. & 82.1 $\pm$ 1.7 & 81.4 $\pm$ 3.7 & 0.74 $\pm$ 0.04 & 0.43 $\pm$ 0.10 & 0.72 $\pm$ 0.04 & 0.46 $\pm$ 0.11\\
\quad + $E_{b}$ FiLM & 82.7 $\pm$ 1.7 & 87.1 $\pm$ 2.0 & 0.71 $\pm$ 0.05 & 0.62 $\pm$ 0.04 & 0.70 $\pm$ 0.04 & 0.66 $\pm$ 0.04\\
\bottomrule
\end{tabular*}
\end{table*}

\section{Results and Discussion}

\subsection{Lineshape Analysis}\label{sec:resspec}

The feasibility of local bond environment classifications directly from the carbon 1$s$ Auger spectral lineshape in organic molecules is first examined. Figure \ref{fig:spectra} shows the class averaged spectra (blue) and individual carbon spectra (grey) for the 13 merged environment classes. The un-merged environment classes spectra are shown in SM Figure S1. A common, broadly defined, lineshape between all the class averaged spectra is prevalent upon visual inspection. All classes show a peak maximum around 250-260 eV, with a lower kinetic energy shoulder or satellite peak about 240 eV. For the aromatic carbons in the heteroaromatic, aryl N, aryl O, aryl F and aryl carbonyl merged classes (top row), this shoulder peak is less well defined than in other classes, and these class spectra are dominated by a single peak. Though the aryl N, aryl O and aryl F classes, show a trend in increasing sub-peak resolution on the left slope of the dominant peak about 240 eV, with increasing atomic number. This observation is unlikely to be an artifact of the sample size in each merged class, as the classes have counts of 214, 247 and 217 respectively. Due to the approximate nature of the calculations used to compute the spectra and the omission of any final-state specific analysis in this work, drawing further conclusions between the specific lineshape deviations and environment classes from visual inspection alone will be avoided. However, it is clear that each merged class environment exhibits unique global and local lineshape features, which encourages the capability of the CNN classification.

To provide a more quantitative analysis of the variation between the Auger spectra lineshapes and the merged environment classes, Figure \ref{fig:pcalda} shows scatter plots of the principal component analysis (PCA) (left) and linear discriminant analysis (LDA) (right) dimensionality reduction techniques on the m carbon spectra. a) colors the scatter point by 1 eV-wide $E_b$ bins and b) colors and shapes the scatter points by the merged environment classes. 

\begin{figure*}[!htbp]
    \centering
    \includegraphics[width=1.0\textwidth]{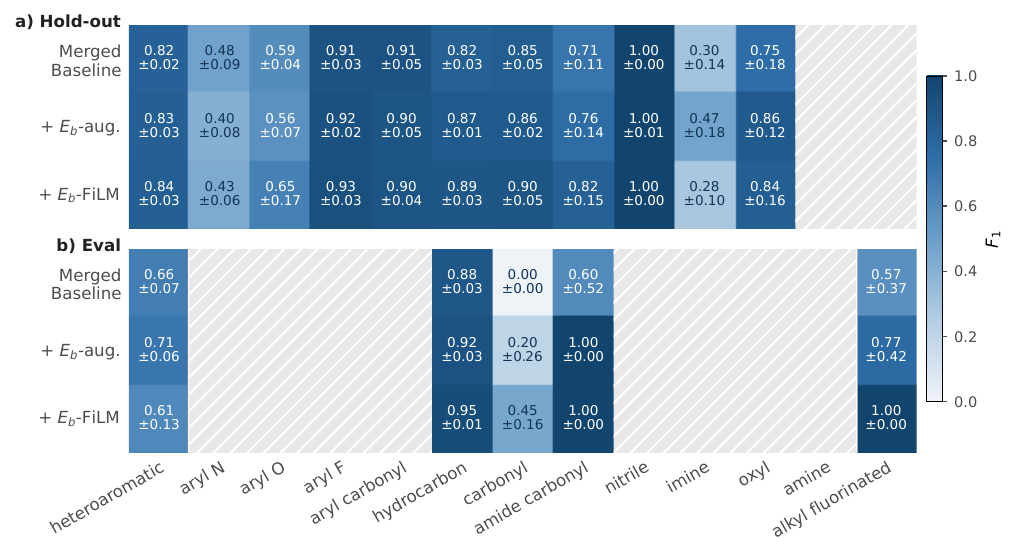}
    \caption{Per-class F$_{1}$ scores on the merged classes with the AES baseline, the $E_b$-aug and $E_b$-FiLM models for the classes in the a) hold-out and b) evaluation test sets. Classes which are not present in either dataset have blank columns.}
    \label{fig:matrix}
\end{figure*}

\begin{figure}[!htbp]
    \centering
    \includegraphics[width=1.0\columnwidth]{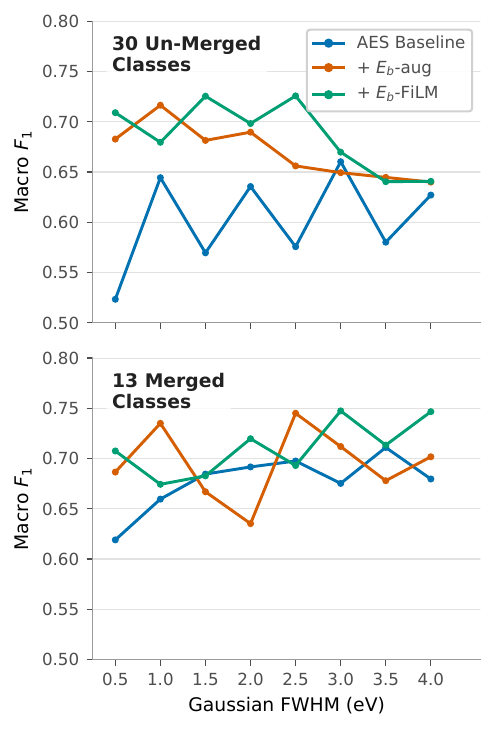}
    \caption{The effect of the Gaussian broadening FWHM on the classification performance on the hold-out test, for the 30 un-merged (top) and 13 merged classes (bottom). The baseline model (AES) is compared to the input lineshape augmented with $E_b$ ($E_b$-aug, dark orange) and the $E_b$ conditioning the classification with FiLM layers ($E_b$-FiLM, green).}
    \label{fig:fwhm}
\end{figure}

PCA is a dimensionality reduction technique that represents data in a basis of principal components (PC), which are the eigenvectors of the covariance matrix of the mean-centered lineshapes. The PCs are a set of ordered vectors representing orthogonal directions in the space of kinetic energies which maximize the variance of lineshape in the dataset. The first two principal components (PC1 and PC2) are shown, and due to the unsupervised nature of PCA a) and b) display the same shape of scatter points with different color schemes. PC1 and PC2 only show 44.7$\%$ of the total explained variance determined by the PCA, which highlights the complexity of the spectra, furthermore the overall degree of clustering between between both categories is weak with a multitude of overlapping regions. The lineshape variance will be influenced by other factors such as the molecular size (Figure \ref{fig:data}) and atoms type composition. These factors which will affect the number of valence electrons which can contribute to the spectrum, and thus density of states and broadness of the spectrum. 

The weak clustering indicates regions where similar binding energies and common bond environments share similar lineshape signatures and indicates the potential for the combined lineshape and binding energies to be used as inputs for a deep learning classifier of the local bond environment. For example, the region of dark blue points around PC1 -2 and PC2 -2 in a) corresponds to a cluster of hydrocarbon environments in b) (dark orange circles). Such environments are known to have relatively low core-electron binding energies as the absence of electronegative neighbors results in a relatively higher electron densities about these carbons, which screens the core-hole more effectively.

The plots on the right show the LDA supervised dimensionality reduction technique. This considers the binding energy and environment classes in a) and b) respectively. LD1 and LD2 show two directions in the kinetic energy space which best separate the classes, by maximizing the difference between the class means and minimizing the spread within each class. In b), the environment classes are grouped in clusters but not cleanly separated, with only the nitriles standing fully apart. In a) the binding energy separates only the highest and lowest values, with little distinction between roughly 292 and 297 eV, as different local environments in different molecules often share a similar binding energies. Therefore, binding energy alone would be insufficient for the classification of local carbon environments, but it reduces the space of possible environment classes and therefore its combination with the lineshape should improve the classification performance, as demonstrated by the following subsection.

\subsection{Classification Performance}\label{sec:resclass}

The classification performance of the un-merged and merged environment class models on the hold-out and evaluation test sets is given in Table \ref{tab:cnn_cv}. The baseline models only consider the Auger spectrum lineshape using a conventional CNN architecture which is inspired by previous work on XPS measurement analysis\cite{PIELSTICKER2023341433}. The $E_b$-aug models include the core-electron binding energy ($E_b$) of the carbon by trivially appending the z-normalized $\Delta E_b$ to the normalized fitted Auger intensities and the $E_b$-FiLM model conditions the CNN feature extraction through the inclusion of FiLM layers between the convolution layers (see Subsection \ref{sec:model}). The results are averaged over the 10 CV folds and the $\pm$ standard deviations are provided. Three performance metrics are reported. Accuracy is the ratio of correct classifications to total classifications. Macro recall is the unweighted class average of the ratio of correct positives to all actual positives, and so penalizes false negatives; for the imbalanced class distribution of Figure \ref{fig:data} it is more informative than accuracy. Macro F$_{1}$ is the unweighted class average of the per-class F$_{1}$ scores, which are the harmonic mean of that class's precision and recall. Precision is the ratio of correct positives to everything assigned that class. Recall improves as false negatives fall and precision as false positives fall, so macro F$_{1}$ balances the two and, like macro recall, represents performance on an imbalanced dataset better than accuracy does.

Generally the performance on the hold-out dataset is better than evaluation dataset, which reflects the difference in the molecular size distributions between the QM9 subset and the evaluation molecules (Figure \ref{fig:data}). The evaluation dataset scores reflect the model transferability to the smaller systems appearing in the organic molecule AES literature. Considering just the AES baseline models first, the accuracy results shows that merging the classes into broader definitions significantly improves model performance with the hold-out and evaluation accuracies going from 76.6$\%$ and 43.3$\%$ to 80.2$\%$ and 76.4$\%$ respectively. However, the macro recall and F$_{1}$ scores show more modest improvements or no improvement at all. For example the hold-out macro recall goes from 0.70 to 0.69 and macro F$_{1}$ goes from 0.67 to 0.68. The macro F$_{1}$ scores show some improved transferability to the evaluation molecules with respect to the class merging with un-merged and merged scores of 0.32 and 0.37. The merged class definitions were produced by a chemical intuition based on similarities in the atom and bond types in the nearest neighbors. It is feasible that better definitions could be derived from a combination of empirical factors from the shape of the spectra and the chemical similarities of the groups.

Table \ref{tab:cnn_cv} clearly shows that including $E_b$ as an additional spectral modality improves the classification performance. The most notable performance increase is seen for the $E_b$-FiLM macro F$_{1}$ scores on evaluation dataset with the merged class definitions, which improve on the AES baseline macro F$_1$ from 0.37 to 0.66, whilst the $E_b$-aug only improves this to 0.46. This indicates that conditioning the CNN model training with $E_b$ provides a more robust approach to multi-modal classification tasks in spectroscopy than trivial augmentation of the inputs. 

For the hold-out datasets the improvements with respect to the inclusion of $E_b$ are more modest. The macro F$_{1}$ increases from 0.67 to 0.68 for $E_b$-aug and to 0.70 for $E_b$-FiLM with the un-merged classes. For the merged classes, the macro F$_{1}$ goes from 0.68 to 0.72 for $E_b$-aug and to 0.70 for $E_b$-FiLM. The reduced performance of $E_b$-FiLM relative to $E_b$-aug for the merged classes can be examined via the per-class F$_{1}$ scores in Figure \ref{fig:matrix}. This unpacks the macro F$_{1}$ scores in Table \ref{tab:cnn_cv} and shows the CV averaged per-class F$_{1}$ scores for the three models for the hold-out and evaluation datasets in a) and b) respectively. $E_b$-FiLM reduces the classification performance on the imine class with respect to both the AES baseline and the $E_b$-aug models. However, it worth acknowledging that there are only 4 imine carbons in the hold-out dataset. Out of the 11 classes present in the hold-out set, $E_b$-aug improves the performance on 7 of the classes with respect to the AES baseline and $E_b$-FiLM improves 7 classes. For the evaluation molecules, the large increase in performance of the macro F$_{1}$ for the $E_b$-aug and  $E_b$-FiLM models can be attributed to the performance on the carbonyl groups which are 0.20 and 0.45 respectively. The AES baseline fails to correctly classify any of the 3 carbonyl carbons in the evaluation dataset. Besides the evaluation heteroaromatic group, and hold-out imine and oxyl groups the $E_b$-FiLM model outperforms the $E_b$-aug model, which further suggests that FiLM layers are a more robust approach for including the $E_b$ as an additional modality to the classification. A likely explanation is that the FiLM layers apply a learned affine transformation equally to every point of the spectral feature maps during training, and therefore exploit the correlation between the binding energy, the environment class and the lineshape shown in Figure \ref{fig:pcalda} more effectively than the augmentation, which merely increases the length of the input from 751 to 752.

Finally, Figure \ref{fig:fwhm} looks at the dependence on the classification performance, via the macro  F$_{1}$, with respect to the Gaussian FWHM applied to the spectra for the 30 un-merged classes (top) and 13 merged classes (bottom). The models are relatively stable to the choice of the FWHM and the performance generally oscillates with respect to the FWHM. This is likely a result of the final adaptive average pooling step of the CNN feature extractor. Adaptive pooling adjusts the size of the pool input automatically to a fixed output size. As an output size of 32 is used in this work and the feature maps received by the pool are the same length as the energy grid that the spectra are fit to, 751 points between 200 and 275 eV. The effective energy grid spacing of the pooled output is 2.3 eV, this demonstrates how pooling in CNNs improves the robust of the models classification with respect to global adjustments to the spectral structure whilst preserving the complexities of the hierarchical patterns captured in the feature maps which enable the classification performance.

\section{Conclusion}

In the present work, I have demonstrated that the wealth of information encoded into AES, resulting from the complexity of the Auger-Meitner decay process, can be potentially exploited with CNNs to perform site-selective organic functional group classification directly from the Auger spectrum lineshape. Furthermore, I have shown that the classification performance is increased when the core-electron binding energy of the atom site undergoing Auger-Meitner decay, is included as an additional modality in the classification. Out of the two approaches explored, conditioning the classification with the binding energy with FiLM layers gives an improved performance over simple augmentation of the input spectra with the binding energy. The success of the FiLM layers showcases how a machine-learning method developed for incorporating text to image classification can be transferred to multi-modal classification in molecular spectroscopy. 

Directly classifying functional groups from the Auger spectrum lineshape and core-electron binding energies demonstrate the potential for data-driven machine-learning approaches to unlock new characterization capabilities on existing spectroscopic techniques which are information rich but difficult to interpret. Whilst the present work only performs the classification on calculated spectra, the demonstrated stability with respect to the FWHM which results from the CNN adaptive pooling, is promising for future model developments applied to experimental spectra. In-order for the potential of this ML driven classification capability to be realized, high resolution AES experiments yielding per-carbon Auger-spectra in a variety of organic molecules would provide a high-quality evaluation dataset for future model assessment. Whilst deep learning methods, including CNNs, are considered to be black boxes, Rieger \textit{et al.} demonstrated that a small CNN model enabled the construction of a latent matrix for interpretable insights about which spectral regions and peak structures related to different functional groups in the classification of IR spectra.\cite{10.1039/d3dd00203a} Future studies could also apply this design architectural design choice and analysis to future AES classification models

\begin{acknowledgments}
I am grateful for the support from the Eric and Wendy Schmidt AI in Science Postdoctoral Fellowship, a Schmidt Futures Program. 
\end{acknowledgments}

\section*{Data and Software Availability Statement}
The data and the software are freely available at https://doi.org/10.5281/zenodo.22285217 and https://doi.org/10.5281/zenodo.22283453 respectively. The Git repository for the software can be found at https://github.com/afouda11/AugerNet.

\section*{Supplementary Material}
The Supplementary Material (SM) for this work contains a table of the carbon environment class SMARTS string patterns and priority scores, and the class averaged and individual carbon spectra for the 30 un-merged environment classes.

\section*{References}
\bibliography{literature}
\end{document}